\documentclass[twocolumn,english,showpacs,amsfonts,amssymb,amsmath,aps,prl]{revtex4}
\usepackage[T1]{fontenc}
\usepackage[latin9]{inputenc}
\usepackage{graphicx}
\usepackage{amssymb}

\usepackage{hyperref}
\usepackage{dcolumn}
\usepackage{bm}

\newcommand{\figwidth}{0.36\textwidth}
\newcommand{\Vg}{V$_\mathrm{G}~$}
\newcommand{\Tc}{T$_\mathrm{c}~$}

\newcommand{\YBCO}{YBa$_{2}$Cu$_{3}$O$_{7-x}~$}
\newcommand{\nH}{$n_{\mathrm{H}}~$}

\begin{document}

\title{Indications of an Electronic Phase Transition in 2D \YBCO Induced
by Electrostatic Doping}

\author{Xiang Leng}

\affiliation{School of Physics and Astronomy, University of Minnesota, Minneapolis,
Minnesota 55455, USA}

\author{Javier Garcia-Barriocanal}

\affiliation{School of Physics and Astronomy, University of Minnesota, Minneapolis,
Minnesota 55455, USA}

\author{Boyi Yang}

\affiliation{School of Physics and Astronomy, University of Minnesota, Minneapolis,
Minnesota 55455, USA}

\author{Yeonbae Lee}

\affiliation{School of Physics and Astronomy, University of Minnesota, Minneapolis,
Minnesota 55455, USA}

\author{A. M. Goldman}

\affiliation{School of Physics and Astronomy, University of Minnesota, Minneapolis,
Minnesota 55455, USA}

\date{\today}
\begin{abstract}
We successfully tuned an underdoped ultrathin \YBCO film into the
overdoped regime by means of electrostatic doping using an ionic
liquid as a dielectric material. This process proved to be
reversible. Transport measurements showed a series of anomalous
features compared to chemically doped bulk samples and a different
two-step doping mechanism for electrostatic doping was revealed.
The normal resistance increased with carrier concentration on the
overdoped side and the high temperature (180 K) Hall number peaked
at a doping level of p$\sim$0.15. These anomalous behaviors
suggest that there is an electronic phase transition in the Fermi
surface around the optimal doping level.
\end{abstract}


\maketitle
The modulation of the electrical charge carrier density in strongly
correlated electron systems using an electric field as an external
control parameter is a long-standing goal in condensed matter physics
due to its potential impact from both fundamental and technological
points of view\cite{AhnRMP,AhnScience1999GBCO}. Among those interesting
materials, the application of field effect concepts to high temperature
superconductors (HTSCs) is an active research field since it could
provide a tool to control the density of the superconducting condensate
in a reversible way, while keeping a fixed structure and avoiding
changing the disorder associated with conventional doping by chemical
substitution\cite{ImadaRMPMI,MannhartSSTreview}. Apart from potential
applications, electrostatic doping of HTSCs is also a useful tool
to study fundamental questions that still remain open such as the
properties of the superconductor-insulator (SI) transition and the
role of quantum criticality \cite{BollingerLSCO,lengYBCOPRL}.

Recently, the development of electronic double layer transistors
(EDLT), that use ionic liquids (ILs) as gate dielectrics, has been
successfully employed to achieve levels of doping of
$10^{15}/cm^{2}$\cite{YeNature,YeonbaeSTO,ShimotaniZnO}. Taking
advantage of such large charge transfers, this technique has been
employed to successfully tune the SI transition in
La$_{2-x}$Sr$_{x}$CuO$_{4}$(LSCO)\cite{BollingerLSCO} and in \YBCO
(YBCO)\cite{lengYBCOPRL}, and to study the low carrier
concentration side of the phase diagram. Both works reveal
interesting physics on the nature of the quantum phase transition
separating the superconducting and insulating phases when the
density of charge carriers is depleted.

Nonetheless, the possibility of accumulating charge carriers by
the field effect to tune an underdoped cuprate into the overdoped
regime has remained elusive. For the 123 family of cuprate
compounds, it has been shown that the CuO$_{2}$ planes are only
indirectly affected by the electric field since the injected holes
are mainly induced in the CuO$_{x}$
chains\cite{SalluzzoPRLNBCOCuOxchain}. Moreover, the systematic
preparation of YBCO bulk samples by tuning the oxygen
stoichiometry in the overdoped region is difficult to achieve and
the highest level of overdoping is limited by the oxygen
stoichiometric concentration, 0.194 holes/Cu for
YBa$_2$Cu$_3$O$_7$\cite{LiangPRB2006}.

Here we present a transport study of an electrostatically doped YBCO
ultrathin film. We exploit the high local electric fields (10$^{9}$
V/m) induced by an ionic liquid at the surface of the sample to tune
the concentration of holes in the superconducting condensate across
the top of the superconducting dome. The experiment reveals that the
electrostatic doping of YBCO involves a different doping mechanism,
as well as an anomalous normal resistance behavior in the overdoped
regime. Surprisingly, we also find that the Hall number measured at
180 K displays a maximum around the optimal doping level that suggests
the occurrence of an electronic phase transition separating the underdoped
and overdoped regimes.

Ultrathin YBCO films were grown by means of a high pressure oxygen
sputtering technique on (001) oriented SrTiO$_{3}$ substrates. The
technical details of sample preparation and characterization can be
found in Ref.~\cite{lengYBCOPRL}. In order to determine the thickness
threshold that separates insulating and superconducting samples, i.e.
the thickness of the dead layer, we produced a series of thin films
with thicknesses ranging from 5 to 10 unit cells. Using standard four-probe
techniques we characterized the R(T) curves for each sample and we
found that the 5-6 unit cells thick samples are insulating and the
superconducting properties are completely recovered in a 7 unit cell
thick film. The EDLT devices were fabricated using films of this thickness,
following the procedure described in Ref.~\cite{YeonbaeSTO}.

\begin{figure}[htb]
\begin{center}
 \includegraphics[width=\figwidth]{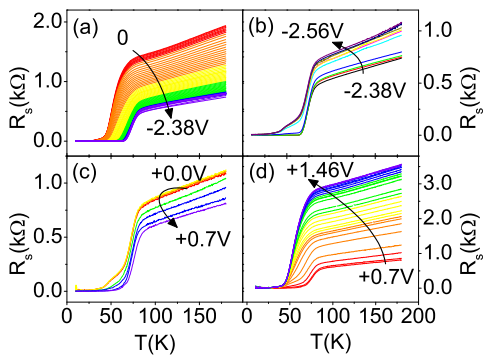}
\end{center}

\caption{(color online) Temperature dependence of the sheet
resistance at different gate voltages. (a). From 0 to -1.2 V, \Vg
was changed in -0.1 V increments and then from -1.2 to -2.38 V,
the increment was -0.02 V. (b). From -2.38 to -2.56 V, \Vg was
changed in increments of -0.02 V. (c). Negative \Vg was then
removed and a positive \Vg was added. From 0 to 0.7 V, \Vg was
changed in 0.1 V increments. (d).\Vg=0.7, 0.75, 0.80, 0.85, 0.90,
0.95, 1.0, 1.05, 1.07, 1.10, 1.13, 1.15, 1.17, 1.20 V, then the
increment was changed to 0.02 V until \Vg=1.46V.
\label{SampleARTVg}}
\end{figure}

We monitored the gating process of the 7 unit cell YBCO ultrathin
film by measuring the sheet resistance as a function of
temperature(see Fig.~\ref{SampleARTVg}). As can be seen in
Fig.~\ref{SampleARTVg}(a), the fresh sample, without any applied
gate voltage (V$_\mathrm{G}$), has a transition temperature \Tc =
42 K. Here \Tc was determined as the temperature corresponding to
the crossing of the extrapolation of the fastest falling part of
the R(T) curve to zero resistance. After the application of a
negative gate voltage, \Tc increases and the normal state sheet
resistance (the metallic region of the curve) drops. This is an
indication that charge carriers are being injected into the sample
and the concentration of holes is increasing towards the optimal
doping point. For higher negative gating, \Vg= -2.24 to -2.38 V,
the rate of change of \Tc and the normal resistance slows down and
saturates. \Tc reaches a maximum value of 67 K and the normal
resistance reaches a minimum of 750 $\Omega$ at 180 K. We also
observed small fluctuations in both \Tc and the normal resistance
in this regime. As shown in Fig.~\ref{SampleARTVg}(b), upon a
further increase of the negative gate voltage, \Tc decreases,
suggesting that the sample has been tuned into the overdoped
regime. In addition, the normal resistance increases with
increasing negative gate voltage, which is different from what is
found in chemically doped bulk samples\cite{NaqibPhysicaC}. For
the highest negative gate voltages (-2.46 V $<$ \Vg $<$ -2.56 V)
we note that the superconducting transition is not abrupt and a
second transition is turned on at the lowest temperature part of
the R(T) curve. We believe that this phenomenon is associated with
the complete accumulation of holes in the top 1 to 2 unit cells
which are the active layer of the sample, with the insulating dead
layer underneath being affected by the high electric field.

To check the reversibility of the process and rule out a possible
chemical reaction on the sample's surface we reversed the polarity
of the gating. The result is shown in Fig.~\ref{SampleARTVg}(c).
After a threshold voltage of about +0.3 V, \Tc starts to increase
and the normal resistance starts to drop, both of which are evidence
of the recovery of optimal doping. With further increase in the positive
gate voltage, shown in Fig.~\ref{SampleARTVg}(d), \Tc drops and
the normal resistance increases, revealing the complete recovery of
the initial state of the sample and confirming that the electrostatic
doping process is reversible.

\begin{figure}[htb]

\begin{center}
 \includegraphics[width=\figwidth]{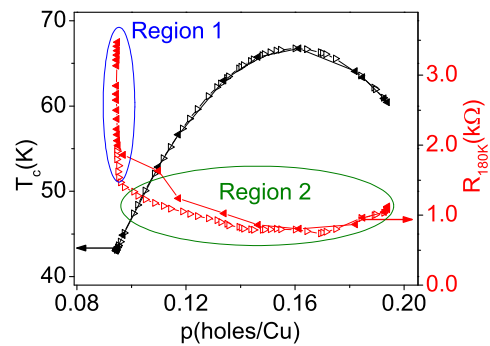}
 \end{center}

\caption{(color online)Transition temperature (\Tc) and normal resistance
at T=180K vs. effective hole doping, p. For both \Tc and the normal
resistance, data with negative values \Vg are denoted by open triangles
and data with positive \Vg, by solid triangles. \label{SampleARTx}}
\end{figure}

In order to obtain an independent variable to describe the process
of accumulation or depletion of holes we inferred an effective
hole doping ``p\textquotedblright{}~ by using the generic
parabolic relation
$T_{\mathrm{c}}/T_{\mathrm{c,max}}=1-82.6(p-0.16)^{2}$\cite{TallonGenericTc}.
We used this calculated number of holes to plot \Tc and the values
of normal resistance at 180 K (see Fig.~\ref{SampleARTx}). The
open triangles represent the evolution of \Tc and normal
resistance at 180 K during the accumulation of holes, the
depletion process is represented by the solid triangles.
Interestingly, the normal resistance at 180 K exhibits a minimum
around the optimal doping point, and its evolution with the number
of holes per CuO$_{2}$ plane roughly follows the same path in both
the depletion and accumulation processes. This is a surprising
behavior since in the overdoped region of the general bulk phase
diagram of cuprates, the normal state is a Fermi liquid and a
lowering of the normal resistance would be expected as the doping
level increased. This raises the possibility that electrostatic
doping is different from the chemical doping of bulk samples. Due
to the short Thomas-Fermi screening length, the electric field
affects only the surface layer (1 to 2 unit cells). Thus we are
studying the properties of a 2D system whose physics might be
different from that of bulk samples. Another possible effect, is
the strong localization of the carriers resulting from the high
local electric field induced by the accumulation of anions of the
ionic liquid at the surface of the sample, or from the eventual
increase of the electronic disorder at high levels of doping.

Upon a closer inspection of Fig.~\ref{SampleARTx}, we notice that
two different regions can be identified during the gating
processes. The first one (region 1) shows an unchanged \Tc over a
large range of normal resistance change (i.e. gate voltages) and
it plays a primary role at the beginning and the end of the
charging processes as holes are accumulated and depleted,
respectively. The second region (region 2) is described above and
it shows that the superconducting \Tc and the normal resistance at
180 K are correlated and the superconducting properties of the
sample are directly affected by the presence of the gate voltage.
This is evidence that a threshold gate voltage is required for the
doping of the superconducting planes. For gate voltages lower than
the threshold, \Tc is not affected and the doping process is
manifested only by the changes of the normal resistance (region
1). Once the threshold is reached the superconducting properties
are affected correspondingly (region 2).

Since the normal resistance is determined by the carriers both on
the CuO$_{2}$ planes and on the CuO$_{x}$ chains while the
superconducting property is mainly determined by the carriers on
the CuO$_{2}$ planes\cite{YBCOphase}, our data suggest that the
electrostatic doping of YBCO is actually a two-step process. First
carriers are induced on the CuO$_{x}$ chains and second, once a
threshold concentration is reached, charge transfer occurs from
the CuO$_{x}$ chains to the CuO$_{2}$ planes. In a previous study
on NdBa$_{2}$Cu$_{3}$O$_{7}$ in a backside gate
configuration\cite{SalluzzoPRLNBCOCuOxchain}, Salluzzo \textit{et
al}. found that the holes injected by the electric field mainly
dope the CuO$_{x}$ chains while the CuO$_{2}$ planes are only
affected by the intra unit cell charge transfer from the CuO$_{x}$
chains. Here we see that this charge transfer can happen only
after a threshold carrier concentration on the CuO$_{x}$ chains is
reached and before this threshold, the electrostatic doping can
affect only the CuO$_{x}$ chains while the CuO$_{2}$ planes are
kept almost untouched. This two-step electrostatic doping process
appears to be different from that of chemical doping, in which any
changes of the hole doping in the range of 0.05$<$p$<$0.2 will
affect the normal resistance and the superconducting properties
simultaneously .

\begin{figure}[htb]

\begin{center}
 \includegraphics[width=\figwidth]{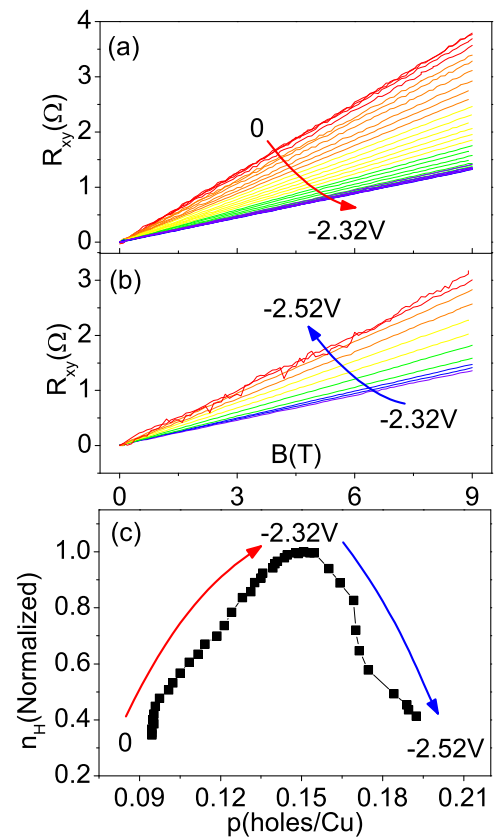}
\end{center}

\caption{(color online)(a) and (b), Hall resistance
(R$_{\mathrm{xy}}$) vs. magnetic field at T=180K for different
gate voltages. From 0 to -2.32 V, \Vg changes -0.04 V every time
and from -2.32 to -2.52 V \Vg changes -0.02 V every time. (c),
Normalized Hall number (n$_{\mathrm{H}}$) vs. hole doping p.
\label{Hall}}
\end{figure}

To develop insight into the nature of the anomalous normal resistance
behavior in the overdoped region, we measured the Hall resistance
at 180 K at different gate voltages. Figures~\ref{Hall}(a) and ~\ref{Hall}(b)
show the linear dependence of the transverse Hall resistance on magnetic
field for different gate voltages. Hall resistances show a positive
slope, which indicates that the charge carriers are holes as expected.
In Fig.~\ref{Hall}(c) the normalized Hall number as a function of
hole concentration is shown. The Hall number is calculated from $n_{\mathrm{H}}=1/(R_{\mathrm{H}}\mathrm{e})$,
being R$_{\mathrm{H}}$ the Hall coefficient which is determined from
the slope of the linear fit of the data plotted in Figs.~\ref{Hall}(a)
and \ref{Hall}(b). Interestingly, we find that the Hall number increases
with the hole concentration up to p$\thicksim$0.15 whereupon an abrupt
change takes place and a peak is developed.

The Hall effect of high-\Tc cuprates has been long studied and two
scenarios have been proposed. One is the two relaxation time
model\cite{AndersonHall1991,ChienHall1991}, in which the Hall
relaxation rate $1/\tau_{\mathrm{H}}~$ and the transport
scattering rate $1/\tau_{\mathrm{tr}}~$ are associate with spins
and charges respectively. In the other scenario one has to
consider the scattering from different regions of the Fermi
surface\cite{StojkoviHall1996}. Experimentally, anomalous
dependence of the normal state Hall coefficient on the doping
level has been reported in chemically doped YBCO\cite{Ando60K},
Bi$_{2}$Sr$_{1.51}$La$_{0.49}$CuO$_{6+\delta}$ (BSLCO)
\cite{BalakirevHallNature} and LSCO\cite{BalakirevPRL2009}. In
YBCO the Hall conductivity at 125 K showed anomalous behavior
around the 60-K phase and in BSLCO and LSCO, $n_{\mathrm{H}~}$
peaks around the optimal doping level at low temperature when
superconductivity is suppressed by ultra high magnetic field.
These anomalous behaviors are closely related to an unusual
electronic state or a sudden change in the Fermi surface, although
chemical doping might also change the magnetic coupling thus
change the Hall scattering rate.

For electrostatic doping, it is hard to imagine that an electric field
will change the magnetic coupling. Thus the \nH peak we observed
indicates that there might be an electronic phase transition in the
Fermi surface. Interestingly, in our results, the peak of the Hall
number is found to be at p$\sim$0.15, to the left of the optimal
doping point (p=0.16). For a doping level of p$=$0.15 the value expected
for T$^{*}$, the pseudogap crossover temperature, roughly matches
180 K \cite{TallonQCP1999}. This would indicate that the electronic
phase transition occurs at the boundary of the pseudogap state and
the strange-metal state in the bulk phase diagram. However, the parabolic
relation we used to derive the effective doping level might not be
quantitatively right for a 2D film. Thus it is only safe to say the
electronic phase transition we observed is around the optimal doping
level, separating the underdoped regime and the overdoped regime.

Quantum oscillation
experiments\cite{YBCOphase,JaudetPRL2008,VignolleNature2008,LeBoeufNature2007}
showed that while there is a large hole-like Fermi surface in the
overdoped regime, there are only small pockets in the underdoped
regime. Angle Resolved Photoemission Spectroscopy also showed that
the transition from the overdoped to the underdoped regime is
accompanied by an electronic reconstruction of the Fermi
surface\cite{NormanARPES1998,ShenScience2005,PlatePRL2005,HossainNature,NormanPhysics2010}.
Thus an electronic phase transition in the Fermi surface is
expected around the optimal doping regime. However, this anomalous
behavior has never been widely reported in bulk samples and the
similar peak observed in Ref.~\cite{BalakirevHallNature}
disappeared at high temperature. This suggest that the electronic
phase transition has been enhanced in our film, either due to the
low dimensionality or the high local electric field.

In summary, we have studied the transport properties of an ultrathin
YBCO film with the doping level changing from the underdoped regime
to the overdoped regime by means of electrostatic doping. Our results
reveal a two-step doping mechanism for the electrostatic doping of
YBCO which is different from that of conventional chemical doping.
Anomalous behavior of the normal resistance on the overdoped side
was also accompanied by a peak in the high temperature (180 K) Hall
number at p$\sim$0.15. This suggests there is an electronic phase
transition on the Fermi surface near the optimal doping point. The
low dimensionality of the film or the high local electric field in
the EDLT configuration might enhance this electronic phase transition
and bring about these anomalous behaviors.

We would like to thank S. Bose and C. Leighton for their
assistance with sample preparation and C. Geppert for his help
with measurements. This work was supported by the National Science
Foundation under grants NSF/DMR-0709584 and 0854752. Part of this
work was carried out at the University of Minnesota
Characterization Facility, a member of the NSF-funded Materials
Research Facilities Network via the MRSEC program, and the
Nanofabrication Center which receives partial support from the NSF
through the NNIN program. JGB thanks the Spanish Ministry of
Education for the financial support through the National Program
of Mobility of Human Resources (2008-2011).

\end{document}